\documentstyle[11pt,epsfig,here,mine]{article}
\begin{document}

\begin{center}
{\Large {\it
Submitted to the Proceedings of 5th International Workshop on\\
TOPICS IN ASTROPARTICLE AND UNDERGROUND PHYSICS\\
(Laboratori Nazionali del Gran Sasso, Assergi,\\
September 7 -- 11, 1997)\\
}}
\end{center}

\vspace{5mm}
\title{STATUS OF THE LAKE BAIKAL EXPERIMENT}

{\vspace{-6mm}
\author{
{\large THE BAIKAL COLLABORATION:}\\[2mm]
V.A.Balkanov$^2$, I.A.Belolaptikov$^7$, L.B.Bezrukov$^1$, N.M.Budnev$^2$, 
A.G.Chensky$^2$, I.A.Danilchenko$^1$, Zh.-A.M.Djilkibaev$^1$, 
G.V.Domogatsky$^1$, A.A.Doroshenko$^1$, S.V.Fialkovsky$^4$, O.N.Gaponenko$^2$, 
A.A.Garus$^1$, T.I.Gress$^2$, A.M.Klabukov$^1$, A.I.Klimov$^6$, 
S.I.Klimushin$^1$, A.P.Koshechkin$^1$, V.F.Kulepov$^4$, L.A.Kuzmichev$^3$, 
S.V.Lovzov$^2$, B.K.Lubsandorzhiev$^1$, M.B.Milenin$^4$, R.R.Mirgazov$^2$, 
A.V.Moroz$^2$, N.I.Moseiko$^3$, S.A.Nikiforov$^2$, E.A.Osipova$^3$, 
A.I.Panfilov$^1$, Yu.V.Parfenov$^2$, A.A.Pavlov$^2$, D.P.Petukhov$^1$, 
P.G.Pokhil$^1$, P.A.Pokolev$^2$, E.G.Popova$^3$, M.I.Rozanov$^5$, 
V.Yu.Rubzov$^2$, I.A.Sokalski$^1$, Ch.Spiering$^8$, O.Streicher$^8$, 
B.A.Tarashansky$^2$, T.Thon$^8$, R.Wischnewski$^8$, I.V.Yashin$^3$
}

\address{
1 - Institute  for  Nuclear  Research,  Russian  Academy  of   Sciences
(Moscow); \mbox{2 - Irkutsk} State University (Irkutsk); \mbox{3 - Moscow}
State University (Moscow); \mbox{4 - Nizhni}  Novgorod  State  Technical
University  (Nizhni   Novgorod); \mbox{5 - St.Petersburg} State  Marine
Technical  University  (St.Petersburg); \mbox{6 - Kurchatov} Institute
(Moscow); \mbox{7 - Joint} Institute for Nuclear Research (Dubna);
\mbox{8 - DESY} Institute for High Energy Physics (Zeuthen) 
}

\maketitle\abstracts{
We review the present status of the Baikal Underwater Neutrino Experiment and 
report on neutrino events recorded with the detector stages  {\it NT-36} and 
{\it NT-96}.
}

\section{Detector and Site}

The Baikal Neutrino Telescope is being deployed in Lake Baikal, Siberia, 
\mbox{3.6 km} from shore at a depth of \mbox{1.1 km}. At this depth, the 
maximum light absorbtion length 470 and \mbox{500 nm} is about 20 m.
Scattering is strongly forward peaked ($\langle \cos \theta \rangle) 
\approx 0.95)$, with a scattering length about 15 m.

{\it NT-200}, the medium-term goal of the collaboration \cite{APP}, will be 
finished in April 1998 and consists of 192 optical modules (OMs) -- see fig.1. 
An umbrella-like frame carries  8 strings, each with 24 pairwise arranged OMs.
Three underwater electrical cables connect the detector with the shore 
station. Deployment of all detector components is carried out  during 5--7 
weeks in late winter when the lake is covered by thick ice.

In April 1993, the first part of {\it NT-200}, the detector {\it NT-36} with 
36 OMs at 3  strings, was put into operation and took data up to March 1995. A
72-OM array, {\it \mbox{NT-72}}, run in \mbox{1995-96}. In 1996 it was 
replaced by the four-string array {\it NT-96}. Summed over 700 days effective 
life time, $3.2\cdot 10^{8}$ muon events have been collected with
\mbox{{\it NT-36, -72, -96}}. Since \mbox{April 6,} 1997, {\it NT-144}, a 
six-string array with 144 OMs, is taking data.

The OMs are grouped in pairs along the strings. They contain 37-cm diameter 
{\it QUASAR} PMTs which have been developed specially for our project. The two
PMTs of a pair are switched in coincidence in order to suppress background
from bioluminescence and PMT noise. A pair defines a {\it channel}. 

\vspace{-1mm}
\begin{figure}[H]
\centering
  \mbox{\epsfig{file=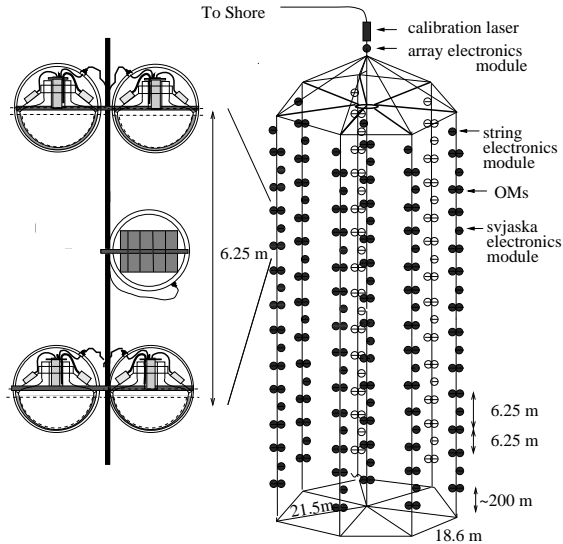,height=7.5cm,angle=-90}}
\vspace{0cm}
\caption[4]{\small
Schematic view of the Baikal Telescope {\it NT-200}.
                  The modules of {\it NT-144}, operating since April
                  1997, are in black.
                  The expansion left-hand shows 2 pairs of
                  optical modules ("svjaska") with the svjaska
                  electronics module, which houses
                  parts of the read-out and control electronics.
}
\end{figure}

\vspace{0mm}
A {\it muon-trigger} is formed by the requirement of \mbox{$\geq N$ {\it hits}}
(with {\it hit} referring to a channel) within \mbox{500 ns}. $N$ is typically
set to \mbox{3 or 4.} For  such  events, amplitude and time of all fired
channels are digitized and sent to shore. A separate {\em monopole trigger} 
system searches for clusters of sequential hits in individual channels which 
are characteristic for the passage of slowly moving, bright objects like GUT 
monopoles.

In the initial project of {\it NT-200}, the optical modules were directed 
alternately upward and downward (fig.1). Due to sedimentation of biomatter 
deterioiating the sensitivity of upward looking OMs we were forced to direct
the OMs of the present arrays essentially downward.

\section{Separation of Neutrino Events}

The signature of neutrino induced events is a muon crossing the detector from 
below. With the flux of downward muons exceeding that of upward muons from 
atmospheric neutrino interactions by about 6 orders of magnitude, a careful 
reconstruction is of prime importance. Two nearly vertical  neutrino events 
have been separated with the rather small {\it NT-36} \cite{ourneu}.
Considering them as atmospheric neutrino events, a 90 \% CL upper limit of 
$1.3 \cdot 10^{-13}$ (muons cm$^{-2}$ sec$^{-1}$) in a cone with 15 degree 
half-aperture around the opposite zenith is obtained (threshold energy $E_{th}
\approx 6$ GeV) with respect to neutrinos due to neutralino annihilation in 
the center of the Earth. 

In contrast to {\it NT-36}, {\it NT-96} can be considered as a real neutrino 
telescope for a wide region in zenith angle $\theta$. After the reconstruction
of all events with $\ge$ 9 hits at $\ge$ 3 strings (trigger{\it 9/3}), quality
cuts have been applied in order to reject fake events. Furthermore, in order 
to guarantee a minimum lever arm for track fitting, events with a projection 
of the most distant channels on the track ($Z_{dist}$) less than 35 meters 
have been rejected. Due to the small transversal dimensions of {\it NT-96}, 
this cut excludes zenith angles close to the horizon (see fig.2).

\vspace{-7mm}
\begin{figure}[H]
\centering
  \mbox{\epsfig{file=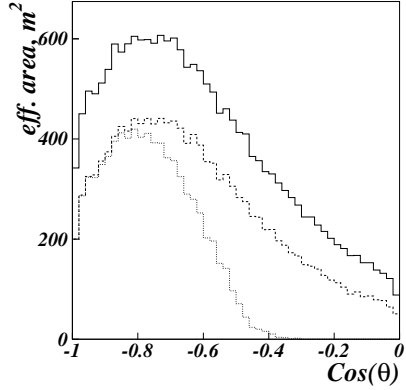,height=5.8cm}}
\vspace{-2mm}
\caption[12]{\small
 Effective area for upward muons satisfying trigger {\it 9/3};
solid line -- no quality cuts; dashed line -- final quality cuts;
dotted line
-- final quality cuts and restriction on $Z_{dist}$ (see text).
}
\end{figure}

\vspace{-0.4cm}
The efficiency of the procedure has been tested with $ 1.8 \cdot 10^6$ MC 
generated atmospheric muons, and with upward muons due to atmospheric 
neutrinos. It turns out that $S/N > 1$ for the lowest curve in fig.2. The 
reconstructed angular distribution of $5.3 \cdot 10^6$ events taken with
{\it NT-96} in April/May 1996 -- after all cuts -- is shown in fig.3

\vspace{-7mm}

\begin{figure}[H]
\centering
  \mbox{\epsfig{file=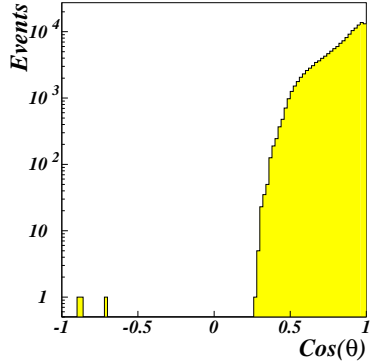,height=5.4cm}}
\vspace{-2mm}
\caption[12]{\small
Experimental angular distribution of events satisfying
trigger
{\it 9/3}, all final quality cuts and the limit on $Z_{dist}$ (see
text).
}
\end{figure}

\vspace{-5mm}

From the first 18 days
lifetime, three neutrino candidates have been separated, in good agreement 
with the expected number of approximately 2.3. Fig.4 displays one of the 
neutrino candidates. Top right the times of the hit channels are shown as a 
function of the vertical position of the channel. At each string we observe 
the time dependence characteristically for upward moving particles. The angle 
regions ${\psi^{min} -\psi^{max}}$ consistent with the observed time 
differences $\Delta t_{ij}$  between two channels {\it i}, {\it j} are given by

\begin{equation}
\label{eq:thetalimit}
\cos(\psi^{min}+\eta) < \cos\psi \frac{c \cdot \Delta t_{ij}}
{\vec{r}_j-\vec{r}_i} < \cos({\psi^{max}-\eta})
\end{equation}

\noindent
with ${\vec{r}_i,\vec{r}_j}$ being the coordinates of the two channels,
$\psi$ the muon angle with respect to $\vec{r}_j-\vec{r}_i$ and $\eta$ the 
Cherenkov angle. The bottom right picture of fig.4 shows that the overlap
region of all channel combinations of this event clearly lay below horizon.
The same holds for  the other two events.

\vspace{-12mm}

\begin{figure}[H]
\centering
  \mbox{\epsfig{file=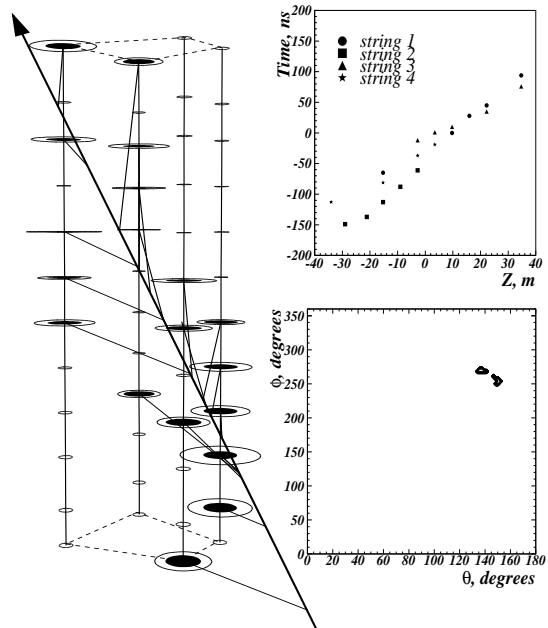,height=10.2cm}}
\vspace{-1.5cm}
\caption[12]{\small
A "gold plated" 19-hit neutrino event. {\it Left:} Event display. Hit channels
are in black. The thick line gives the reconstructed muon path, thin lines 
pointing to the channels mark the path of the Cherenkov photons as given by 
the fit to the measured times. The areas of the circles are proportional to 
the measured amplitudes. {\it Top right:} Hit times versus vertical channel 
positions. {\it Bottom right:}  The allowed $\theta/\phi$ regions (see text) 
The fake probability of this event is smaller than 1\%.
}
\end{figure}

\vspace{-3mm}

In the mean time, altogether 70 days from {\it \mbox{NT-96}} have been 
analyzed, and 12 neutrino candidates have been found. Nine of them have been
fully reconstructed, 3 nearly upward vertical tracks hit only 2 strings and 
give a clear zenith angle but ambiguities in the azimuth angle -- similar to 
the two events from {\it NT-36}. Taking into account the degradation of 
{\it NT-96} due to failed OMs, this is in agreement with MC expectations.

\section{Outlook}

The Baikal detector is well understood, and first atmospheric neutrinos have
been identified. Also muon spectra have been measured, and limits on the 
fluxes of magnetic monopoles as well as of neutrinos from WIMP annihilation in
the center of the Earth have been derived. 

The {\it NT-200} detector will be completed in April 1998. In the following 
years, it will be operated as a neutrino telescope with an effective area 
between 1000 and 5000 m$^2$ typically, depending on the energy. It will 
investigate atmospheric neutrino spectra above 10 GeV (about 1 atmospheric 
neutrino per day). Presumably still too small to detect neutrinos from AGN and
other extraterrestrial sources, {\it NT-200} can be used to push the flux 
limits for neutrinos from WIMP annihilation and for magnetic monopoles. It 
will also be a unique environmental laboratory to study water processes in 
Lake Baikal.

Apart from its own value, {\it NT-200} is regarded to be a prototype, for the 
development a telescope 20-50 times larger. With 2000 OMs, a threshold of 
10-20 GeV and an effective area of 50,000 to 100,000 m$^2$, this telescope 
would have a realistic detection potential for extraterrestrial sources of 
high energy neutrinos. With its comparatively low threshold, it would fill a 
gap between underground detectors and  planned high threshold detectors of 
cube kilometer size.

\vspace{2cm}
{\it This work was supported by the Russian Ministry of Research,the German 
Ministry of Education and Research and the Russian Fund of Fundamental 
Research ( grants }{\sf 96-02-17308} {\it and} {\sf 97-02-31010}).

\end{document}